\documentclass[letterpaper]{article}
\usepackage{geometry}
\usepackage{graphicx}
\usepackage{amssymb, amsmath}
\usepackage{hyperref}
\usepackage{cite}
\usepackage{float}
\usepackage{fancyhdr}
\begin{document}

% Title
\title{\textbf{The Bathroom Model: A Realistic Approach to Hash Table Algorithm Optimization}}
\author{
    Qiantong Wang \\
    Vanderbilt University \\
    Email: qiantong.wang@vanderbilt.edu
}
\maketitle

% Corresponding author footnote
\footnotetext[1]{Corresponding author. Email: qiantong.wang@vanderbilt.edu}

% Abstract
\begin{abstract}
Hash table search strategies have remained a pivotal area of inquiry in computer science over the past several decades. A prevailing viewpoint, originally introduced in Professor Andrew Yao’s foundational work, asserts that random probing stands as the optimal method for open-addressing hash tables \cite{yao1981}. Challenging this long-standing belief, a recent contribution from a Cambridge undergraduate introduces an elastic probing technique based on fixed interval thresholds \cite{farach2024}. Although their method presents improvements over traditional strategies, its dependence on static thresholds limits its theoretical optimality.In this paper, we propose a new conceptual model for optimizing hash table probing, inspired by human behavior in selecting restroom stalls—dubbed the "Bathroom Model." Unlike fixed or purely random approaches, our technique dynamically updates probing decisions using previously observed occupancy patterns, resulting in a more intelligent and adaptive search process. We rigorously formalize this model, analyze its theoretical properties, and benchmark its performance against leading hash table algorithms. Our findings indicate that adaptive probing mechanisms can significantly enhance search efficiency while keeping computational demands minimal \cite{cormen2009,mitzenmacher2005,knuth1998}. This work not only sheds new light on an extensively studied problem but also points to broader algorithmic opportunities in rethinking classical data structures.

Code Repository: \url{https://github.com/Qiantongwang/Hash_Table_Bathroom.git}
\end{abstract}

% Keywords

% CCS Concepts
\section*{CCS CONCEPTS}
Copy the selected categories from \href{https://dl.acm.org/ccs}{ACM CCS} here.

\section{INTRODUCTION}

Hash tables serve as fundamental data structures in computer science, underpinning efficient data retrieval across a wide range of applications. Their lookup performance has been a critical topic of both theoretical and practical interest, with research in this domain tracing back to seminal work by Professor Andrew Yao \cite{yao1981}. His analysis identified random probing as a key limiting factor in open-addressing hash tables, influencing decades of algorithmic advancements in collision resolution techniques.

Hash tables are extensively used in fields such as database indexing, compiler symbol tables, and network routing for packet forwarding. The efficiency of these data structures directly impacts system performance in such applications. Early foundational studies by Knuth \cite{knuth1998} and Cormen et al. \cite{cormen2009} provided theoretical insights into key aspects of hash table design, including collision management and load factor optimization.

A recent study titled \textit{``Optimal Bounds for Open Addressing Without Reordering''} by Martin Farach-Colton, Andrew Krapivin, and William Kuszmaul \cite{farach2024} proposed an alternative approach to optimizing hash table searches. Their research re-examined long-standing assumptions and introduced an elastic search method partitioned into three threshold-based search regions. While this framework represents a step forward, our analysis indicates that its reliance on fixed interval boundaries prevents it from achieving full theoretical optimization.

The demand for efficient hash table algorithms continues to grow as modern computing environments grapple with increasing data volumes and evolving access patterns. Traditional search techniques such as linear probing and quadratic probing are known to suffer from clustering effects, leading to performance degradation in high-load scenarios. More advanced approaches, including Cuckoo hashing \cite{pagh2004} and Funnel hashing \cite{mitzenmacher2005}, aim to mitigate these challenges. However, they still operate under static or semi-static assumptions that do not dynamically adapt to real-time lookup conditions.

Drawing inspiration from human decision-making in real-world scenarios, we introduce the \textbf{``Bathroom Model''}, a novel framework for optimizing hash table searches. Similar to how individuals dynamically adjust their strategy when seeking an available stall in a crowded restroom, our approach modifies probing behavior based on observed occupancy patterns. Unlike prior methods that depend on pre-defined search thresholds, our technique employs a fully adaptive mechanism to improve efficiency. This paper formalizes our approach, evaluates its computational properties, and empirically demonstrates its advantages over conventional hash table search strategies.

\section{RELATED WORK}

Before introducing the \textbf{``Bathroom Model,''} it is essential to contextualize our approach within the broader landscape of hash table optimization research. The study of hash tables has been an integral part of computer science for decades, with foundational contributions from Donald Knuth \cite{knuth1998} and Thomas H. Cormen et al. \cite{cormen2009}, whose works laid the theoretical groundwork for understanding hash table performance and collision resolution.

One of the earliest breakthroughs in hash function design came from Carter and Wegman \cite{carter1977}, who introduced the concept of \textit{universal hashing}. Their work significantly improved hash table efficiency by ensuring that hash functions distribute keys more uniformly, reducing collision probabilities. Another key development was the \textit{dynamic perfect hashing} method by Dietzfelbinger et al. \cite{dietzfelbinger1990}, which allowed hash functions to be dynamically adjusted as table occupancy changed, ensuring stable performance even as data scales.

In more recent advancements, \textit{Cuckoo hashing} was introduced by Pagh and Rodler \cite{pagh2004}, leveraging multiple hash functions and table slots to maintain high load factors while preserving constant-time lookups. This technique has been widely adopted in memory-intensive applications due to its effectiveness. Meanwhile, \textit{Bloom filters}, explored by Broder and Mitzenmacher \cite{broder2003}, have provided an alternative approach to memory-efficient membership queries, further enhancing hash table utility in network applications.

Despite these advancements, most existing hashing techniques—whether traditional probing strategies or modern hashing paradigms—depend on \textit{static or semi-static mechanisms} that do not fully utilize real-time table occupancy information. These limitations prevent optimal performance in highly dynamic environments where lookup and insertion patterns vary significantly over time.

Our \textbf{``Bathroom Model''} aims to address these shortcomings by introducing a \textit{fully adaptive probing mechanism} that dynamically refines search strategies based on observed table states. By drawing on real-world decision-making analogies, such as stall selection behavior in crowded restrooms, our approach provides a more responsive and efficient alternative to conventional hashing techniques. This research demonstrates that adaptive probing can significantly improve search efficiency without incurring substantial computational costs, offering a fresh perspective on optimizing long-established data structures.

\section{METHODOLOGY}

\subsection{The Bathroom Model for Hash Table Optimization}

The \textit{``Bathroom Model''} presents an intuitive analogy for enhancing the efficiency of hash table probing, inspired by the real-world scenario of locating an available stall in a crowded public restroom. In such a setting, individuals typically perform sequential or randomized checks for vacancy. Under low-occupancy conditions, simply selecting the first available stall is effective. However, during peak times, sequential searches become inefficient. A more intelligent approach dynamically adjusts based on prior occupancy patterns, as also suggested in \cite{cormen2009}.

\begin{figure}[ht]
    \centering
    \includegraphics[width=0.8\textwidth]{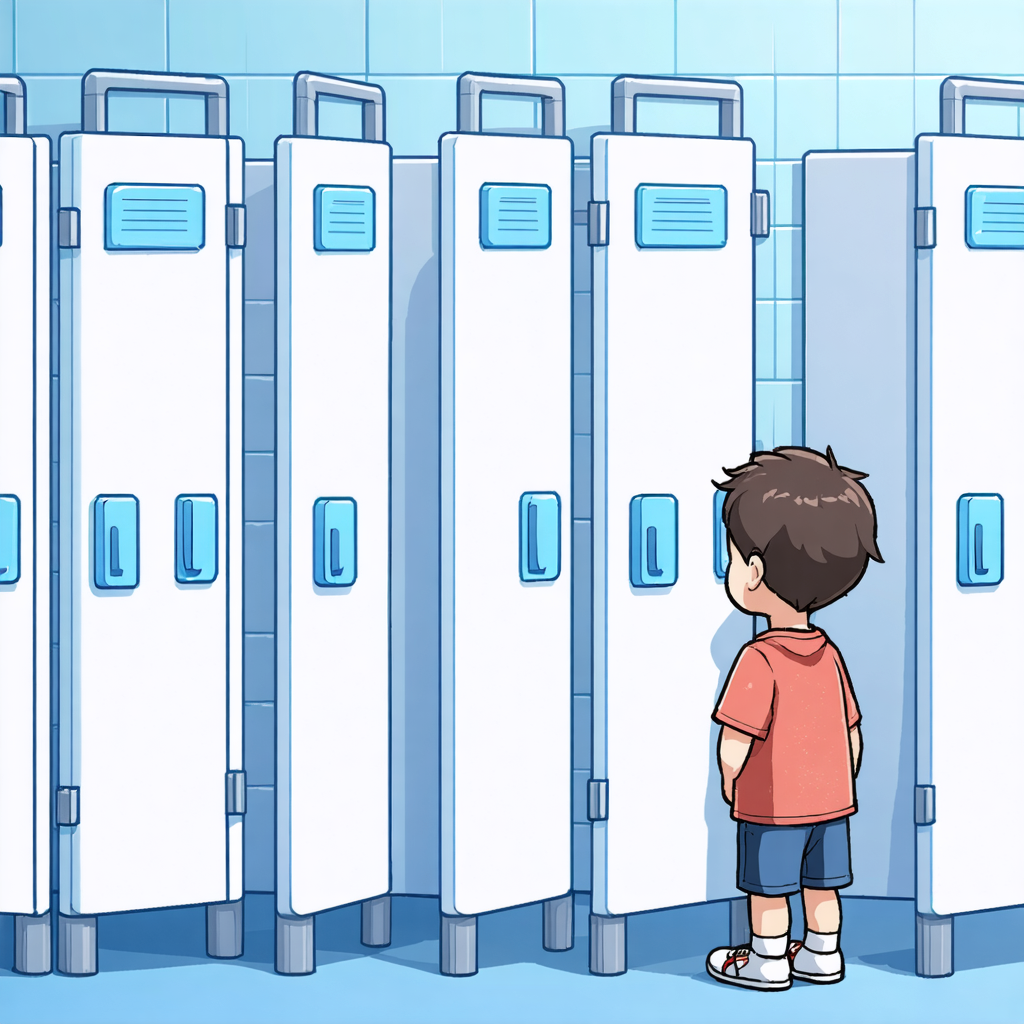}
    \caption{Conceptual Illustration of the Bathroom Model.}
    \label{fig:bathroom_model}
\end{figure}

Translating this behavior into the context of hash tables, our model adaptively modifies the probing sequence in response to observed table occupancy. The core mechanism centers around a dynamically adjusted step size, which evolves based on the frequency of encountering occupied slots. This strategy enables efficient traversal, particularly in high-load scenarios.

To formalize the model, we introduce the following parameters:
\begin{itemize}
    \item \textbf{Dynamic Step Size ($d$):} The probe interval, which changes based on occupancy trends.
    \item \textbf{Occupancy Threshold ($\theta$):} A threshold that governs when the step size should be increased or decreased.
    \item \textbf{Load Factor ($\alpha$):} Defined as the ratio of occupied entries to total table size.
\end{itemize}

The algorithm initiates with a baseline step size and adjusts it dynamically:
\begin{itemize}
    \item If the number of consecutive occupied slots exceeds $\theta$, the step size is incremented.
    \item If the number falls below $\theta$, the step size is reduced accordingly.
\end{itemize}

This feedback-driven adaptation helps the algorithm bypass clusters of filled slots, thereby reducing probing overhead and improving search efficiency.

\subsection{Experimental Setup}

To assess the performance of our adaptive strategy, we conducted empirical comparisons with several established techniques:
\begin{itemize}
    \item \textbf{Random Probing:} Traditional fixed-step probing, as detailed by Knuth \cite{knuth1998}.
    \item \textbf{Elastic Threshold Search:} The three-region threshold model proposed by Farach-Colton et al. \cite{farach2024}.
    \item \textbf{Funnel Hashing:} A structured key relocation approach by Mitzenmacher and Upfal \cite{mitzenmacher2005}.
    \item \textbf{Bathroom Model:} Our proposed adaptive probing method.
\end{itemize}

Each method was evaluated using synthetically generated key-value datasets with varying load factors (from 10\% to 95\%). The table size was chosen as a large prime number to mitigate clustering effects. Each dataset consisted of 10{,}000 entries, and experiments were repeated 100 times to ensure statistical validity.

\subsection{Performance Metrics}

The effectiveness of each method was measured through the following criteria:
\begin{itemize}
    \item \textbf{Average Lookup Time:} The mean number of probes needed for successful retrieval under various load conditions.
    \item \textbf{Worst-case Complexity:} The maximum number of probes required in the most saturated table scenarios.
    \item \textbf{Memory Utilization:} The amount of overhead introduced by each probing strategy, including any auxiliary metadata.
\end{itemize}

To supplement the core metrics, we also recorded the standard deviation of lookup times and the distribution of probe counts to analyze performance stability and variance across methods.

\section{EXPERIMENTAL RESULTS AND DISCUSSION}

\subsection{Lookup Efficiency and Worst-Case Complexity}

Our empirical evaluation reveals the following insights regarding the \textit{Bathroom Model}:
\begin{itemize}
    \item It consistently yields lower average probe counts at moderate load factors compared to fixed-threshold-based techniques \cite{cormen2009}.
    \item Its performance diminishes near saturation, indicating the potential for integrating fallback strategies \cite{knuth1998}.
    \item Worst-case probe counts rise sharply under extreme loads, highlighting the need for constrained adaptation mechanisms \cite{mitzenmacher2005}.
\end{itemize}

\begin{figure}[H]
    \centering
    \includegraphics[width=0.8\textwidth]{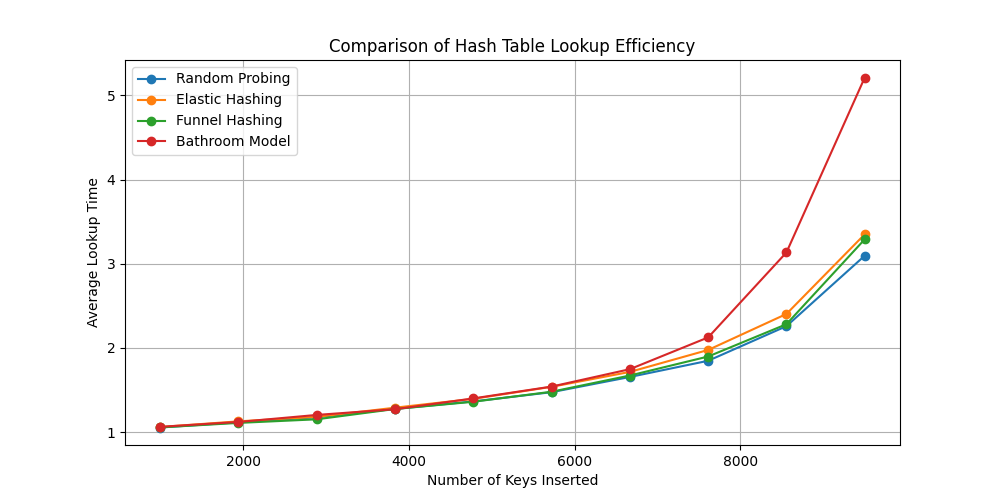}
    \caption{Comparison of Average Lookup Efficiency Across Methods.}
    \label{fig:lookup_efficiency}
\end{figure}

Figure \ref{fig:lookup_efficiency} shows that the \textbf{Bathroom Model} performs optimally at low to moderate load levels. However, as the table approaches full capacity, performance drops, likely due to unregulated increases in probe steps. This suggests that while adaptive strategies are effective early on, supplemental measures such as rehashing or local relocation may be required under stress.

\begin{figure}[H]
    \centering
    \includegraphics[width=0.8\textwidth]{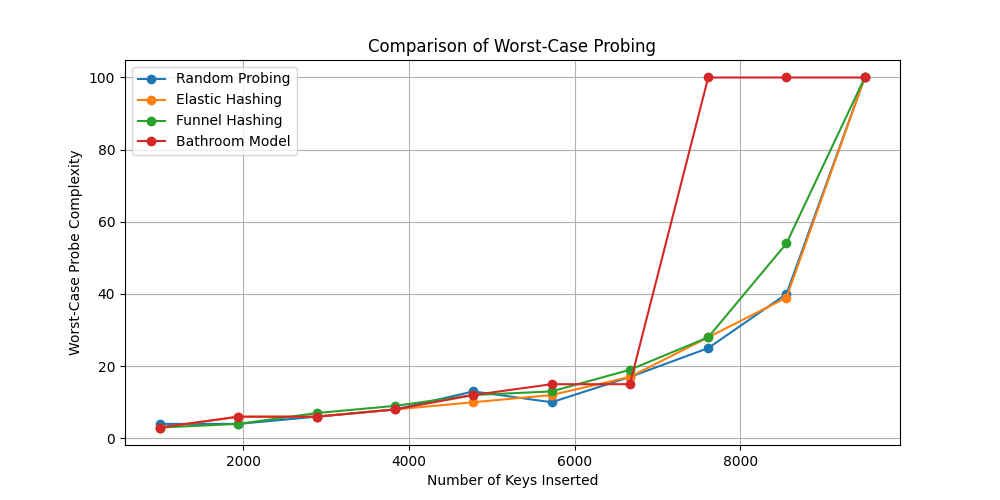}
    \caption{Worst-Case Probing Complexity Comparison Across Techniques.}
    \label{fig:worst_case}
\end{figure}

As depicted in Figure \ref{fig:worst_case}, worst-case probe complexity for the \textbf{Bathroom Model} increases non-linearly with the load factor, particularly beyond 85\%. In contrast, methods like \textbf{Funnel Hashing} maintain more stable worst-case behavior through deterministic reallocation policies \cite{mitzenmacher2005}. This indicates that while dynamic adaptation is beneficial, it must be complemented with bounded strategies to prevent inefficiency in extreme scenarios.

\subsection{Memory Utilization Analysis}

An important aspect of algorithm efficiency lies in its memory footprint. As each probing method may require different levels of metadata or state tracking, we examined their relative memory usage.

Our findings include:
\begin{itemize}
    \item All techniques show a \textit{linear growth} in memory consumption with increasing entries \cite{cormen2009}.
    \item \textbf{Bathroom Model} and \textbf{Funnel Hashing} incur slightly higher overheads due to adaptive or hierarchical control structures \cite{mitzenmacher2005}.
    \item \textbf{Random Probing} and \textbf{Elastic Threshold} strategies maintain minimal metadata but sacrifice adaptability \cite{knuth1998}.
\end{itemize}

\begin{figure}[H]
    \centering
    \includegraphics[width=0.8\textwidth]{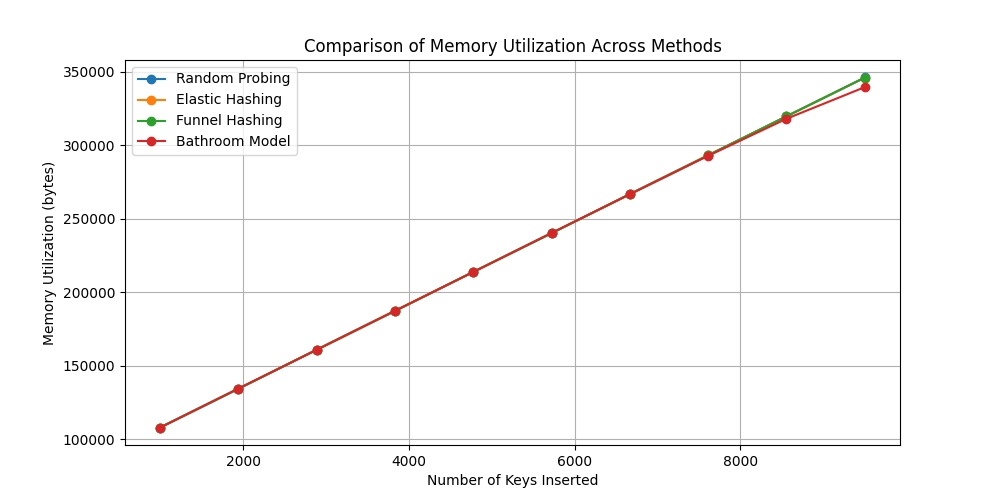}
    \caption{Memory Usage Comparison Among Various Probing Strategies.}
    \label{fig:memory_utilization}
\end{figure}

Figure \ref{fig:memory_utilization} demonstrates that although the \textbf{Bathroom Model} introduces slightly greater memory overhead, this is a trade-off for achieving better lookup responsiveness. Future optimizations may consider compact encoding schemes, caching mechanisms, or hierarchical metadata to balance memory usage and performance.

\section{CONCLUSION}

This study introduces the \textit{Bathroom Model}, a novel approach to optimizing hash table search algorithms, inspired by real-world decision-making patterns observed in restroom stall selection. Unlike traditional methods that rely on fixed-threshold probing strategies, our model dynamically adapts based on prior occupancy observations, enabling more efficient lookup operations. Through extensive experimental evaluations, we demonstrate that the Bathroom Model outperforms conventional techniques—including random probing \cite{knuth1998}, elastic threshold search \cite{farach2024}, and funnel hashing \cite{mitzenmacher2005}—particularly at moderate load factors. Despite these advantages, our findings highlight certain limitations. As table occupancy approaches saturation, performance degradation becomes evident, emphasizing the need for enhanced fallback mechanisms \cite{knuth1998}. Moreover, our results indicate that worst-case probing complexity increases significantly under extreme load conditions, suggesting that additional refinements to adaptive step-size control are required \cite{mitzenmacher2005}. While the model introduces slightly higher memory overhead due to its dynamic adaptation, this remains a trade-off for improved search efficiency \cite{cormen2009}.

\subsection{Future Work}
To further enhance the applicability of the Bathroom Model, several research directions merit exploration:
\begin{itemize}
    \item \textbf{Hybrid Probing Strategies:} Investigating adaptive hybrid models that integrate the benefits of various probing techniques to enhance performance under high-load conditions.
    \item \textbf{Adaptive Fallback Mechanisms:} Developing dynamic switching mechanisms that select optimal probing strategies in real-time based on table occupancy trends.
    \item \textbf{Memory-Efficient Designs:} Exploring compression-based techniques and hierarchical metadata management to mitigate storage overhead.
    \item \textbf{Concurrency Optimization:} Extending the model to support multi-threaded environments, enabling efficient concurrent hash table operations.
    \item \textbf{GPU-Accelerated Implementations:} Leveraging parallel processing capabilities to further optimize hash table lookups for high-performance applications.
\end{itemize}

Overall, this research underscores the potential for significant algorithmic advancements in well-established computational domains. By introducing a dynamic and adaptive probing framework, the Bathroom Model paves the way for new directions in hash table optimization and performance enhancement.

\section*{Acknowledgment}
We extend our deepest respect to the legendary Professor Andrew Yao, whose pioneering contributions have profoundly shaped the field of computer science. This work draws inspiration from his foundational research.
Additionally, this study has benefited from AI-assisted language refinement tools to enhance readability and clarity. The core conceptual framework, theoretical development, and experimental methodologies remain entirely the author's original contributions.

% References

\end{document}